\documentstyle[psfig,conf_iap,]{article}
\begin{document}
\renewcommand{\H}{{\mbox{${\rm H{\sc i}~}$}}}
\newcommand{\Hp}{{\mbox{${\rm H{\sc ii}~}$}}}
\newcommand{\He}{{\mbox{${\rm He{\sc i}~}$}}}
\newcommand{\Hep}{{\mbox{${\rm He{\sc ii}~}$}}}
\newcommand{\Hepp}{{\mbox{${\rm He{\sc iii}~}$}}}
\newcommand{\h} {{\rm H{\sc i}}}
\newcommand{\hp} {{\rm H{\sc ii}}}
\newcommand{\he} {{\rm He{\sc i}}}
\newcommand{\hep} {{\rm He{\sc ii}}}
\newcommand{\hepp} {{\rm He{\sc iii}}}
\newcommand{\NHI}{\mbox{$N_{\rm H{\sc i}}$}}
\newcommand{\mnras}{MNRAS}
\newcommand{\apjl}{ApJL}
\newcommand{\pasp}{PASP}
\newcommand{\aap}{A\&A}
\renewcommand{\thefootnote}{\fnsymbol{footnote}} 
\heading{%
%
Detecting \Hep reionization from a sudden injection of entropy in the
intergalactic medium\footnote{Based on the data obtained from the OPC
program 65.O-0296A (P.I.  S.~D'Odorico) at the VLT/Kueyen telescope,
ESO, Paranal, Chile.}\\
%
} 
\par\medskip\noindent
\author{%
Tom Theuns$^{1}$, Saleem Zaroubi$^{2}$ and Tae-Sun Kim$^{3}$
}
\address{Institute of Astronomy, Madingley Road, Cambridge CB3 0HA, UK}
\address{Max-Planck Institut f\"ur Astrophysik, Postfach 123, D-85740
Garching bei M\"unchen, Germany}
\address{European Southern Observatory, Karl-Schwarzschild-Stra\ss e 2, D-85748 Garching bei
M\"unchen, Germany}

\begin{abstract}
The temperature of the low-density intergalactic medium is set by the
balance between adiabatic cooling resulting from the expansion of the
universe, and photo-heating by the UV-background. A sudden injection
of entropy from the reionization will increase the temperature of the
gas, leading to a broadening of the hydrogen Lyman-$\alpha$ absorption
lines produced in the IGM, and observed in the spectra of background
quasars. We present a method based on wavelets to characterise
objectively the line widths of such absorption lines. We use high
resolution hydrodynamical simulations to demonstrate that the
algorithm can detect changes in temperature of order of 50 per cent on
scales $\ge 5000$ km s$^{-1}$. We apply the method to a UVES/VLT
spectrum of quasar 0055--269 ($z_{\rm em}=3.7$) and detect at the 99
per cent confidence level a sudden increase in temperature below
redshift $z\sim 3.3$, which we interpret as evidence for \Hep
reionization.
\end{abstract}
\section{Introduction}
The intergalactic medium (IGM) can be observed in the spectra of
distant quasars through resonant absorption in the hydrogen
Lyman-$\alpha$ transition \cite{1}. The absence of strong absorption
throughout the spectrum (Gunn-Peterson trough) indicates that the IGM
is very highly ionised, $\rho_{\rm H\sc{I}}/\rho_{\rm H}\sim 10^{-4}$
below redshifts $z\sim 6$, but the nature of the first ionising
sources remains uncertain.

If the sources responsible for reionizing \H are sufficiently hard,
then \Hep will be reionized at the same time. However, if for example
galaxies were the main contributor to the UV-background, then \Hep
reionization may be significantly delayed until a population of harder
sources appears. The large fluctuations in the \Hep optical depth
observed with STIS and FUSE \cite{2} suggest \Hep reionization occurs
late, around $z\sim 3$. There is supporting evidence from the
hardening of the UV-background, as deduced from metal line ratios
although these results remain controversial \cite{3}.

An independent approach is to study the temperature of the
IGM. Photo-heating and adiabatic expansion introduce a tight
density-temperature relation \cite{4}, $
T=T_0(\rho/\langle\rho\rangle)^{\gamma-1}\,,$ for the low density IGM
responsible for producing the Lyman-$\alpha$ forest. A sudden injection
of entropy resulting from reionization will increase $T_0$ and make the
gas nearly isothermal, $\gamma\sim 1$. Detecting reionization through a
sudden entropy increase has the advantage that one determines the epoch
at which a significant fraction of the volume of the universe is
overrun by the ionization front.

Schaye et al. used hydrodynamical simulations to demonstrate that the
density-temperature relation introduces a cut-off in the scatter plot
of column density $N_\H$ versus line width $b$ \cite{5}. They used a
set of many high resolution simulations to calibrate the relation
between the position of the cut-off and $(T_0,\gamma$). Applying the
calibration to the cut-off measured in ten high-resolution spectra,
they found evidence for a rise in $T_0$ around a redshift $z\sim 3.3$,
and an associated dip in $\gamma$, which they associated with \Hep
reionization. Ricotti, Gnedin \& Shull (\cite{6}) applied a similar
technique, but calibrated with pseudo hydrodynamical simulations, to
published line lists, and found a similar temperature increase,
although their error bars are large and their result is consistent
with a non-evolving $T_0$ as well.  Bryan \& Machacek \cite{7} also
found evidence for a high value of $T_0$, but McDonald et al. \cite{8}
did not find an increase in $T_0$ around $z=3.3$, although they used
largely the same data as Schaye et al. Zaldarriage \cite{9} applied a
wavelet analysis similar to the one discussed here to look for
temperature changes in the spectrum of QSO 1422+231 which might be a
relic from reionization, and constrained them to be smaller than a
factor of 2.5.

\section{Wavelet analysis}
\begin{figure}
\centerline{\vbox{
\setlength{\unitlength}{1cm}
\centering
\begin{picture}(7.,7.)
\put(-2.5, -4.0){\includegraphics{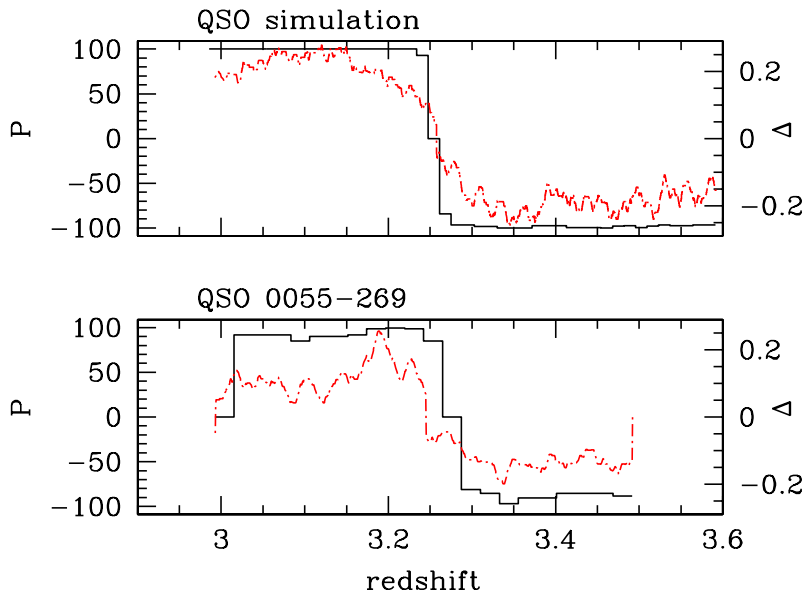}}
\end{picture}
}}
\caption[] {Application of the wavelet method to a simulated spectrum (top
panel) and QSO~0055--269 (bottom panel). In the simulations, the low
redshift half is 50 per cent hotter than the high redshift half. This
jump is detected at the 99.5 per cent level. A similar jump is seen in
the observed spectrum.}
\label{fig:ttfig1}
\end{figure}

A discrete wavelet is a localised function with a finite
bandwidth. This makes wavelets useful for characterising line widths
in a spectrum, since the amplitude of the wavelet will be related to
the width of the line, and the position of the wavelet to the position
of the line. The decomposition is moreover unique, for a given wavelet
basis.

Theuns \& Zaroubi \cite{10} used the Daubechies 20 wavelet to
characterise temperature fluctuations in simulated Lyman-$\alpha$
spectra.  They demonstrated how the wavelet amplitudes are large when
the gas is cold and the lines narrow, and {\em vice versa}. They also
showed how the cumulative distribution of wavelet amplitudes can be
used to characterise the gas temperature, and to judge whether two
stretches of spectrum have different temperatures or not. Here, we
have used $\Delta$, the maximum difference between the cumulative
distribution of a stretch of spectrum, and the distribution for the
spectrum as a whole, as a measure of how much $T_0$ for a stretch,
differs from $\langle T_0\rangle$.

We have extended this method to attach a statistical significance to
differences in cumulative distributions. Given a line list, we
construct mock spectra by scrambling the lines, thereby producing a
spectrum where the lines have the {\em same shapes} as in the original
spectrum, but any {\em correlation} in line widths that might result
from temperature variations along the spectrum has been removed. We
can now attach a statistical level of significance for the correlation
between line widths -- and hence the temperature of the gas -- by
comparing the statistics of the original spectrum with the scrambled
one.


In our simulation, we chose the photo-ionization rates for Hydrogen
and Helium such that \H and \He reionize around redshift $z\sim 7$ but
\Hep reionization is delayed until $z\sim 3.4$. In order to
investigate the effect of different temperatures, we impose a power
law equation of state, $T=T_0\,(\rho/\langle\rho\rangle)^{\gamma-1}$,
on the simulation output before computing mock spectra. We have used
$(T_0,\gamma)=(1.5\times 10^4K,5/3)$ and $(2.2\times
10^4K,5/3)$. These spectra are made to look like real data as much as
possible, by adding photon and read-out noise, and broadening the
lines with the instrumental profile of the spectrograph. Voigt
profiles are fitted to the absorption lines with the same programme as
used to analyse the data.

\section{Results}
Our results are shown in Fig.~(\ref{fig:ttfig1}). The dot--dashed line
(right hand scale) is the maximum difference between the cumulative
distribution of wavelet coefficients for a window of size 8000 km
s$^{-1}$ at redshift $z$, and the spectrum as a whole. Histogram (left
hand scale) is the significance (in \%) for unusually large values of
$|\Delta|$, obtained from comparison with spectra constructed from
randomised line lists. Negative values indicate gas colder than
average, and vice versa for positive values. In the simulated spectrum
(top panel), the first and second halves of the spectrum have
$T_0=2.2\times 10^4$K and $1.5\times 10^4$K respectively. The wavelet
statistic $\Delta$ detects this imposed jump, as can be seen from the
sudden change in the sign of $\Delta$.  The significance of the change
is computed to be $\ge 99.5$ per cent.  The bottom panel applies the
analysis to the spectrum of QSO~0055--269. We detect a sudden change
in the value of $\Delta$ around $z\sim 3.3$, significant at the $\ge
99.5$ per cent. Note that no simulations are used to compute the level
of significance: the method described here is solely based on the line
list.

We note that the high redshift gas is {\em colder} than the lower
redshift gas. This is opposite to what one would expect from
photo-heating in the optically thin limit, in which case the IGM will
{\em smoothly cool down}. Therefore we suggest this sudden temperature
increase is due to \Hep reionization at redshift $\sim 3.3$.

\acknowledgements{We thank S. D'Odorico, S. Cristiani, E. Giallongo, A,
Fontana and S. Savaglio for allowing us to use the UVES data prior to
publication. TT thanks PPARC for the award of a post-doctoral
position.This work has been supported by the \lq Formation and
Evolution of Galaxies\rq\ and \lq Physics of the Intergalactic
Medium\rq\ networks set up by the European Commission.  Research
conducted in cooperation with Silicon Graphics/Cray Research utilising
the Origin 2000 super computer at the Department for Applied
Mathematics and Theoretical Physics, Cambridge.}

\vfill
\end{document}